# Evaluating the effect of topic consideration in identifying communities of rating-based social networks

Ali Reihanian
Department of Information Technology
Mazandaran University of Science and Technology
Babol, Iran
areihanian@ustmb.ac.ir

Behrouz Minaei-Bidgoli
Department of Computer Science
Iran University of Science and Technology
Tehran, Iran
b_minaei@iust.ac.ir

Muhammad Yousefnezhad
Department of Computer Science
Nanjing University of Aeronautics and Astronautics
Nanjing, China
myousefnezhad@nuaa.edu.cn

*Abstract*—Finding meaningful communities in social network has attracted the attentions of many researchers. The community structure of complex networks reveals both their organization and hidden relations among their constituents. Most of the researches in the field of community detection mainly focus on the topological structure of the network without performing any content analysis. Nowadays, real world social networks are containing a vast range of information including shared objects, comments, following information, etc. In recent years, a number of researches have proposed approaches which consider both the contents that are interchanged in the networks and the topological structures of the networks in order to find more meaningful communities. In this research, the effect of topic analysis in finding more meaningful communities in social networking sites in which the users express their feelings toward different objects (like movies) by the means of rating is demonstrated by performing extensive experiments.

## I. INTRODUCTION

With the emergence of social networks, people have been attracted to them, and have been sharing valuable information by means of communicating with each other. For example, folksonomies are social tagging sites which their users collaboratively express their feelings and sentiments toward a special resource like a movie or music by means of descriptive keywords (tags) [1] or ratings. One of the most important issues considered when analyzing these kinds of networks is community detection. A Community (also sometimes referred to as a module or cluster [2]) is a dense sub-network within a larger network, such as a close-knit group of friends in a social network [3]. The community structure of complex networks reveals both their organization and hidden relations among their constituents [4].

A large number of methods have been proposed to extract appropriate communities from networks including a set of nodes (individuals) and weighted edges (connections). They just consider the graph structure of the network for finding communities and no content analysis has been used in the process of their proposed approaches.

Despite of the original definition of the networks, nowadays, real world networks like Facebook and Twitter are containing a vast range of information including shared objects, comments, etc. It is unreasonable for a community to be explained by a single entity because the community members are generally interacting with each other via a large number of distinguishable ways in various domains.

One of the possible solutions is to find topical clusters in which the nodes have the same topic of interest. Each topical cluster represents one of the topics of interest in the network. Then, a community detection algorithm can be applied to these topical clusters to find the ultimate communities [5]. In this way, we can analyze and estimate the effect of topic consideration in community detection.

In this paper, the effect of topic analysis in finding more meaningful communities in social networking sites in which the users express their feelings toward different objects (like movies) by the means of rating, is demonstrated by performing extensive experiments. Therefore, a network is partitioned into different topical clusters in which the nodes have the same topic of interest. Then, a community detection algorithm is applied to the topical clusters in order to find more meaningful communities. This will lead us to communities in which the nodes are tightly connected and have the same topic of interest. This process is called topic-oriented community detection [5]. At last, the results of community detection with topic consideration are compared with the results of community detection without considering the topics of interest. Quantitative evaluations reveal that the results of community detection will be improved when the topic of interest in the network is considered.

The remainder of the paper is outlined as follows. In section II, related works are reviewed. Section III explains the topic-oriented community detection. In order to evaluate the effect of topic consideration in identifying the communities of rating-based social networks, extensive experiments are conducted on real-life data sets. The descriptions of these data sets, the experimental results and their analyses are given in section IV. Finally, the conclusions are given in section V.

## II. RELATED WORKS

Many researches have been done in the area of community detection. Most of these researches mainly focus on the topological structure or linkage patterns of networks. They merely consider the graph structure of the network for finding communities, while no content analysis is used in the process of their proposed approaches.

According to the community detection strategies which were employed in these researches, their proposed methods can be classified into optimization-based methods and heuristic methods. Some of the optimization-based methods focus on optimizing an objective function [5]. One of the most important works in the literature was a research done by Newman and Girvan, in which they introduced modularity as an objective function [6]. A large amount of works has been done to optimize modularity such as the methods which were developed in [7-9]. This function has been influential in the literature of community detection, and has gained success in many applications. Modularity is used to evaluate the quality of a particular division of a network into communities [5]. On the other hand, heuristic methods such as GN algorithm [10] and CPM algorithm [11] design a graph clustering algorithm based on intuitive assumptions [5].

Even though these researches have gained success in some applications, since they mainly focus on the topological structure of the networks, they ignored the contents interchanged between members. As a result, the relationships between the members in these researches are mainly based on the total number of communications.

In recent years, a number of researches have proposed approaches which consider both the contents that are interchanged in the networks and the topological structures of the networks in order to find more meaningful communities. Z. Zhao, et al. proposed a topic-oriented community detection approach based on social objects' clustering and link analysis [5]. Their proposed approach could identify the topical communities which reflect the topics and strengths of connections simultaneously. Zhu, et al. combine classic ideas in topic modeling with a variant of mixed-membership block model which is recently developed in the statistical physics community [12]. In their research, Zhu, et al. combine topic-modeling with link structure. A. Zhao and Ma proposed a framework to apply a semantically structured approach to the Web service community modeling and discovery [13].

## III. TOPIC-ORIENTED COMMUNITY DETECTION IN A SOCIAL NETWORK

As we mentioned earlier, the goal of this paper is to demonstrate the effect of topic consideration in finding more meaningful communities in social networking sites in which the users express their feelings toward different objects by the means of rating. For this cause, some components of the frame work which was proposed in [5] are changed in order to be applicable to the mentioned social networks. This framework detects communities which have unique topic of interest and connected members. Each community contains the nodes of the network which have the same topic of interest. This framework is implemented in four steps: Preprocessing and annotating topic labels, Clustering social objects, Creating topical clusters and Applying a community detection algorithm to the topical clusters.

### A. Preprocessing and annotating topic labels

In this step, data sets are preprocessed and ready to use. In this process, the social objects are recognized. Generally, People communicate with each other through social objects. These objects often imply the topics which people are interested in. Social objects can be classified into two kinds of situations [5]: 1) the social objects which are attached to multi-members, 2) the social objects which are attached to one member.

In the first situation, the edges between members are built because of a social object. An example of this situation can be happened in a movie rating network. In this network, edges between members are built when they rate the same movie. As a matter of fact, in this network, each movie (social object) is attached to multi members. The members of the movie rating network are connected to each other due to the rating of the same movie.

In the second situation, each social object is attached to only one member. Therefore the social objects are considered to be the attributes of members of the network. An example of this situation can be happened in a paper citation network. In this network, papers (members) cite each other. Also, each paper contains a text content (the title of a paper) which is a social object and can be considered as the attribute of the corresponding paper.

Figure 1 shows the two different kinds of relations between the members of a network and social objects. The network which is located in the left side of Figure 1 is a movie rating network. As it is clear, the edges between members are built because of the social objects. Also, the network which is located on the right side of the Figure 1 is a paper citation network. In this network, each social object is the attribute of its corresponding paper.

Since in this paper the social networking sites in which the users express their feelings toward different objects are analyzed, the first situation is happened.

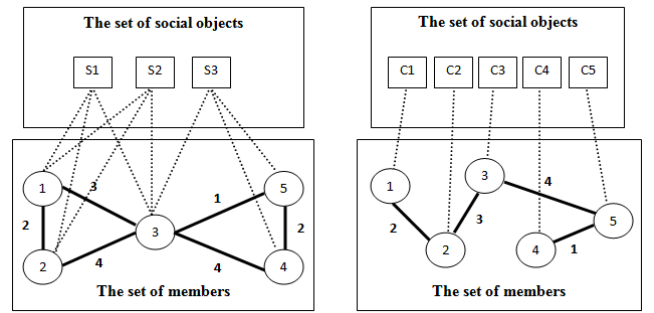

Fig. 1. Two different kinds of relations between the members of networks and social objects

So, in this step, data sets are preprocessed and ready to use. In this process, the social objects are recognized. Afterward, the topics of each social object in the data set are retrieved. Subsequently, each social object is labeled by its corresponding topic. In some cases the topics of each social object can easily be retrieved manually, or there are corresponding tags which represent the topics for each social object. But in cases where a social object is represented by text and its labels cannot easily be retrieved, a method has been introduced by Z. Zhao, et al. which can annotate the topic label to each social object [5].

*B. Clustering social objects*

In this step, social objects in a network are partitioned into different clusters. Each cluster represents a unique topic which is shared by its members. In other words, According to their labeled topics, social objects are partitioned into different clusters in a way that each cluster includes members with the same topic.

Different methods can be used to perform the social object clustering according to the type of social objects. For example, a novel method has been proposed in [5] to cluster the text social objects. This method combines the vector space model with the Entropy Weighting K-Means (EWKM) [14] in order to cluster the text social objects. Since the data sets which are used in this paper contain social objects with labeled topics, we manually partition these social objects into different clusters.

*C. Creating topical clusters*

Using the results that are generated in the previous step, we partition the members of the network into different topical clusters. In the first step, each social object has been annotated with a topic label. In this step, members are partitioned into different topical clusters with considering the topic labels of the social objects they are involved in. Thus in this step we find clusters in which every member has the same topic of interest. Therefore the total number of topical clusters is equal to the number of topics of interest in the network. A user can be a member of several topical clusters, since it is common for a user to be interested in several topics.

*D. Applying a community detection algorithm to the topical clusters*

This step aims to find communities in each of the topical clusters which were created in the previous step. Members in each topical cluster are connected to each other with different strengths. Based on the number of ratings on the same social objects, some members may have stronger connections, while some other may have weak or no connections. This has been concluded according to the topic analysis that has been performed in the framework. Since the result of the framework is to detect communities which have unique topic of interest and connected members, we should apply a community detection algorithm to the previously created topical clusters in order to identify the tightly connected members.

In order to perform this process, many community detection algorithms can be employed such as GN and so on. Newman proposes an important algorithm to partition network graphs of links and nodes into sub graphs. He also introduces a concept which is called modularity. In the case of weighted networks, modularity has been defined as follows [9]:

$$Q = \frac{1}{2m} \sum_{i,j} \left[ A_{ij} - \frac{k_i k_j}{2m} \right] \delta(c_i, c_j) \quad (1)$$

Where $A_{ij}$ represents the weight of the edge between i and j, $k_i = \sum_j A_{ij}$ is the sum of the weights of the edges attached to vertex i, $c_i$ is the community to which the vertex i is assigned, the δ function δ (u, v) is 1 if u=v and 0 otherwise and also $m = \frac{1}{2} \sum_{ij} A_{ij}$.

Since Newman's algorithm was very time-consuming, Blondel, et al. suggest the modified version of the algorithm in order to make it faster, giving rise to what is known as the "Louvain method" [15]. This algorithm is a modularity maximization algorithm which iteratively optimizes the modularity in a local way and aggregates nodes of the same community [16]. In this paper, the "Louvain method" has been applied in order to find topical communities.

IV. EXPERIMENT AND ANALYSIS

In this section, the results of our research are presented. First, three real-life data sets along with a performance metric which were used in the experiments are described. Then, the process of detecting topical communities in the mentioned data sets is discussed and its results are analyzed. Finally, the results of topic-oriented community detection (with performing content analysis) are compared with the results of community detection without performing any content analysis.

*A. Real life data sets*

We used the publicly available data sets in our experiments which are Movielens, Book-Crossing and CIAO datasets. Movielens data set [17] is a rating data set which is collected from the Movielens web site (http://movielens.org). It consists of 100000 ratings from 943 users which were given to 1682 movies. Book-Crossing data set [18] is a rating data set which is collected from the Book-Crossing community (http://www.bookcrossing.com). It contains 278858 users providing 1149780 ratings about 271379 books. CIAO data set [19-22] is a rating data set which is collected from a product review site (http://ciao.com) in which users share their opinions about a product by means of rating or commenting. There are 35773 ratings in this data set which are attached to 16850 products by 2248 users.

As described earlier in this paper, the topic-oriented community detection framework considers the results of topic analysis for finding more meaningful communities. So, in order to evaluate this framework, two aspects should be considered: topic and linkage structure. It means that the expected results should keep each community's members with the same topic and strong connections. Zhao et al. introduced a performance evaluation metric which considers both topic and linkage structure [5]. This metric has been defined as follows:

$$PurQ_\beta = (1 + \beta^2)(Purity \cdot Q)/(\beta^2 \cdot Purity + Q) \quad (2)$$

As it is clear in the above equation, The PurQ$_\beta$ has three parameters which are Q, Purity and β. Q denotes the modularity. This parameter measures the communities from the perspective of the link structure. The larger the Q, the better the communities are divided from the perspective of topological structure. In our experiment, for each topical cluster, modularity is calculated by equation 1. Since the topic-oriented framework may generate more than one topical cluster for each data set, the total value of modularity in this framework is calculated as follows:

$$Q = \sum_{i=1}^{n} \frac{Weight_{TC_i}}{Weight_T} \cdot Q_{TC_i} \qquad (3)$$

Where n is the number of generated topical clusters. $Q_{TC_i}$ is the value of modularity for the topical cluster $TC_i$. $Weight_{TC_i}$ is the sum of the weights of edges in the topical cluster $TC_i$. $Weight_T$ is the sum of the weights of edges in the topical cluster, which is directly created from the basic network (when no topical clustering has been performed). It should be considered that since in this framework no communications' content analysis is performed, the weight of each relationship between two members is the number of ratings which are given to the same social objects by these two members.

In equation 2, Purity represents the purity of topics in the detected communities and is calculated as follows [5]:

$$Purity = 1/N_{cm} \cdot \sum_{i=1}^{N_{cm}} \max_{1 \leq j \leq k} \{n_{ij}/n_i\} \qquad (4)$$

Where $N_{cm}$ represents the number of detected communities, $n_{ij}$ refers to the number of nodes belonging to topic j and community i, $n_i$ refers to the number of nodes in community i. k is the number of topics in the network. The higher the Purity, the better the communities are partitioned from the perspective of topics.

β is a parameter to adjust the weight of Purity and Q and β $\in [0, \infty]$. If we consider the purity of topics and the topology of the network equally important, the value of β should be set to 1. If we want to pay more attention to Purity in comparison with Q, then the value of β should be set to a number between 1 and ∞. On the other hand, if we want to pay more attention to Q in comparison with Purity, the value of β should be set to a number between 0 and 1. Actually β is used in equation 2 to adjust the emphasis of topics and link structure [5].

*B. Experiments*

In order to identify the communities by applying the topic-oriented community detection framework to the three introduced datasets, four steps (according to section III) have been taken. The first step was to preprocess the data sets. As to the Movielens and Book-Crossing data sets, movies and books were considered as the social objects. So, for the Movielens data set the genres of the movies were extracted. These extracted genres are the same as the genres attached to each movie by IMDB (http://www.imdb.com). Then, all the movies which were in the genres of Documentary or Western were retrieved. As you know, the genre of a movie represents the general topic in which a movie is made about. In this step, we achieved 77 movies. For the Book-Crossing data set, we extracted the categories of 93 books from Amazon (http://www.amazon.com). As for the CIAO data set, products were considered as the social objects. Each product's category was attached to it in the data set. Thus for the Book-Crossing data set and the CIAO data set, the categories represent the topics of each product or book.

The second step was to cluster the social objects. As for the Movielens data set, the movies were partitioned into two clusters of Documentary and Western. The Documentary cluster contained 50 movies while the Western one contained 27 movies. As for the Book-Crossing data set, the books were partitioned into two clusters of Fiction and Non-Fiction. The Fiction cluster contained 80 books, while the Non-Fiction cluster contained 13. The products in the CIAO data set were partitioned into six clusters of DVDs, Books, Beauty, Music, Travel, and Food and Drink. The DVDs cluster contains 2057 products, The Books cluster contains 2803 products, the Beauty cluster contains 2333 products, the Music cluster contains 1801 products, the Travel cluster contains 3922 products and finally the Food and Drink cluster contains 3937 products.

The third step was to create topical clusters. Therefore in each data set, the users who rate the social objects in each cluster were partitioned into topical clusters. For example, all users who rate the movies in the cluster of "Documentary" were partitioned into the topical cluster of "Documentary". The members of each topical cluster rated the social objects which have the same topics. Thus according to the number of topics, we achieved two topical clusters for the Movielens and Book-Crossing data sets and 6 topical clusters for the CIAO data set. As mentioned earlier, since in this framework no communications' content analysis is performed, the weight of each relationship between two members is the number of ratings which are given to the same social objects (for example, two movies in the genre of Documentary) by these two members.

The last step was to detect topical communities. Thus we applied the "Louvain method" to each topical cluster created in the previous step. In order to accurately calculate the modularity, we applied the Louvain method to each topical cluster ten times, and calculated the average of the achieved values of modularity.

Table I gives the results achieved by applying the topic-oriented community detection framework to the Movielens, Book-Crossing and CIAO data sets. In this Table, the columns "Topical Clusters", "No. of Edges" and "No. of Nodes" represent the created topical clusters in the process of applying the topic-oriented framework to the three mentioned data sets, the number of edges and the number of nodes existing in each of these topical clusters, respectively. Moreover, the columns "Total Modularity" and "Purity" denote the overall modularity value (Q) and Purity value for all of the topical communities.

TABLE I. THE RESULTS ACHIEVED BY APPLYING THE TOPIC-ORIENTED COMMUNITY DETECTION FRAMEWORK TO MOVIELENS, BOOK-CROSSING AND CIAO DATA SETS

| *Data sets* | *Topical Clusters* | *No. of Edges* | *No. of Nodes* | *Total Modularity* | *Purity* |
|---|---|---|---|---|---|
| *Movielens* | Documentary | 15833 | 352 | 0.1244 | 1 |
| | Western | 69369 | 491 | | |
| *Book-Crossing* | Fiction | 8531 | 1021 | 0.8469 | 1 |
| | Non-Fiction | 1587 | 191 | | |
| *CIAO* | DVDs | 53916 | 1356 | 0.3086 | 1 |
| | Books | 8999 | 904 | | |
| | Beauty | 5267 | 811 | | |
| | Music | 2076 | 569 | | |
| | Travel | 12905 | 867 | | |
| | Food & Drink | 29763 | 1193 | | |

As it is clear in Table I, Purity has its maximum value in each of the three data sets. The reason is that, the topical clusters created in each data set incorporate members which are interested in the same unique topics. Therefore the purity of topics in each of the topical communities is 1 according to equation 4. It should be considered that it is possible for a certain user to be in several topical clusters, since the interest of people in several different topics is common. Thus some of the members of topical clusters in each data set may be the same. For example, consider the case that a user rated several different movies. Some of these movies were in the genre of Documentary, and the others were in the genre of Western. Therefore this user belongs to both topical clusters in the Movielens data set.

*C. Comparison*

In order to prove the superiority of the results of detecting communities with topic consideration, in this section, we compare the results of topic-oriented community detection, which was implemented in section *B*, with the results of Classical Community Detection in which no content analysis is performed.

In the process of classical community detection approach, a community detection algorithm is applied to a network (basic network) in which the weight of the edges represents the number of communications between relevant nodes. In this condition, no content analysis is done.

We first applied the "Louvain method" to the basic networks of the Movielens, Book-Crossing and CIAO data sets (implementing the Classical Community Detection Framework). Then we partitioned the basic networks of the three mentioned data sets into topical clusters. Each topical cluster includes members which have the same topic. Afterwards, the Louvain method was applied to these topical clusters (implementing the Topic-oriented community detection framework which was discussed in section *B*). We then used $PurQ_\beta$ to evaluate the performances in the experimental evaluation. The corresponding results are given in Table II. Consequently, as it is shown in Table II, β was set to 0.5, 0.75, 1, 1.5, 2 respectively, which represents the different strengths for the topic and the link. Purity, Q and $PurQ_\beta$ have been calculated for each of the two mentioned frameworks.

According to Table II, Modularity and Purity has higher values in the topic-oriented framework, since the basic network is partitioned into topical clusters, and each identified community includes members who have the same topic of interest. Therefore, the topic-oriented community detection framework has a higher value of $PurQ_\beta$ for all five values of β.

TABLE II. COMPARISON OF MODULARITIES WHICH WERE ACHIEVED BY APPLYING THE TOPIC-ORIENTED FRAMEWORK ALONG WITH CLASSICAL COMMUNITY DETECTION FRAMEWORK TO EACH OF THE THREE MENTIONED DATA SETS.

| *Data set* | *Frameworks* | *Total Modularity* | *Total Purity* | $PurQ_\beta$ | | | | |
|---|---|---|---|---|---|---|---|---|
| | | | | *β=0.5* | *β=0.75* | *β=1* | *β=1.5* | *β=2* |
| *Movielens* | Classical | 0.1086 | 0.9777 | 0.3760 | 0.2519 | 0.1955 | 0.1495 | 0.1321 |
| | Topic-oriented | 0.1244 | 1 | 0.4154 | 0.2830 | 0.2213 | 0.1703 | 0.1509 |
| *Book-Crossing* | Classical | 0.8375 | 0.9050 | 0.8906 | 0.8795 | 0.8699 | 0.8572 | 0.8502 |
| | Topic-oriented | 0.8469 | 1 | 0.9651 | 0.9389 | 0.9171 | 0.8888 | 0.8737 |
| *CIAO* | Classical | 0.2899 | 0.8279 | 0.6038 | 0.4963 | 0.4294 | 0.3624 | 0.3332 |
| | Topic-oriented | 0.3086 | 1 | 0.6906 | 0.5535 | 0.4716 | 0.3920 | 0.3581 |

V. CONCLUSION

This paper evaluates the effect of topic consideration in finding more meaningful communities in social networking sites in which the users express their feelings toward different objects (like movies) by the means of rating. Therefore, the network is partitioned into different topical clusters in which the nodes have the same topic of interest. Then, a community detection algorithm is applied to the topical clusters in order to detect communities. After that, a comparison has been performed between the results of topic-oriented community detection and the results of Classical Community Detection in which no content analysis is performed. The experimental results indicate that the results of topic-oriented community detection will be improved when it is joined with topic analysis.

There is a plenty of room to study on community detection problem in real complex networks which contain huge amount of information with different natures. Therefore, in future works we have a plan to work on the effect of other kinds of contents in the network, like the communications' content analysis, in finding more meaningful communities in social networking sites in which the users express their feelings toward different objects with rating.